\documentclass[aps,pra,showpacs,letter,showkeys]{revtex4}
\usepackage{amsfonts}
\usepackage{amsmath}
\usepackage{amssymb}
\usepackage{graphicx}
\begin{document}
\title{Cooling in reduced period optical lattices:
non-zero Raman detuning}
\author{V. S. Malinovsky}
\affiliation{MagiQ Technologies Inc., 171 Madison Avenue, Suite
1300, New York, New York 10016}
\author{P. R. Berman}
\affiliation{Michigan Center for Theoretical Physics, FOCUS
Center, and Physics Department, University of Michigan, Ann Arbor,
MI 48109-1120} 

\begin{abstract}
In a previous paper [Phys. Rev. A \textbf{72}, 033415 (2005)], it
was shown that sub-Doppler cooling occurs in a standing-wave Raman
scheme (SWRS) that can lead to reduced period optical lattices.
These calculations are extended to allow for non-zero detuning of
the Raman transitions. New physical phenomena are encountered,
including cooling to non-zero velocities, combinations of Sisyphus
and "corkscrew" polarization cooling, and somewhat unusual origins
of the friction force. The calculations are carried out in a
semi-classical approximation and a dressed state picture is
introduced to aid in the interpretation of the results.
\end{abstract}

\maketitle

\section{Introduction}
In a previous paper \cite{rmn} (hereafter referred to as I), we have shown
that sub-Doppler cooling occurs in a \emph{standing-wave Raman scheme} (SWRS).
The SWRS is particularly interesting since it is an atom-field geometry that
leads to optical lattices having reduced periodicity. Reduced period optical
lattices have potential applications in nanolithography and as efficient
scatterers of soft x-rays. Moreover, they could be used to increase the
density of Bose condensates in a Mott insulator phase when there is exactly
one atom per lattice site. With the decreased separation between lattice
sites, electric and/or magnetic dipole interactions are increased, allowing
one to more easily carry out the entanglement needed in quantum information
applications \cite{der}.

In this paper, the calculations of I, which were restricted to two-photon
resonance of the Raman fields, are extended to allow for non-zero Raman
detunings. There are several reasons to consider non-zero detunings. From a
fundamental physics viewpoint, many new effects arise. For example, one finds
that, for non-zero detuning, it is possible to cool atoms to non-zero
velocities, but only if \textit{both} pairs of Raman fields in the SWRS are
present, despite the fact that the major contribution to the friction force
comes from atoms that are resonant with a \textit{single }pair of fields. This
is a rather surprising result since it is the only case we know of where
non-resonant atoms that act as a catalyst for the cooling. Moreover,
comparable cooling to zero velocity and non-zero velocities can occur
simultaneously, but the cooling mechanisms differ. We also find effects which
are strangely reminiscent of normal Doppler cooling, even though conventional
Doppler cooling is totally neglected in this work. A dressed atom picture is
introduced to simplify the calculations in certain limits; however, \textit{in
contrast to conventional theories of laser cooling}, nonadiabatic coupling
between the dressed states limits the usefulness of this approach. The
non-adiabatic transitions result from the unique potentials that are
encountered in the SWRS. To our knowledge, there are no analogous calculations
of laser cooling in the literature.

From a practical point of view, there is also a need for calculations
involving non-zero detunings. For example, in the quantum computing scheme
proposed in \cite{der}, the Raman frequency differs at different sites owing
to the presence of an inhomogeneous magnetic field, making it impossible to be
in two-photon resonance throughout the sample. As a result, one has to assess
the modifications in cooling (and eventually trapping) resulting from non-zero
detunings.
\begin{figure}[t]
\centerline{\scalebox{0.4}{\includegraphics{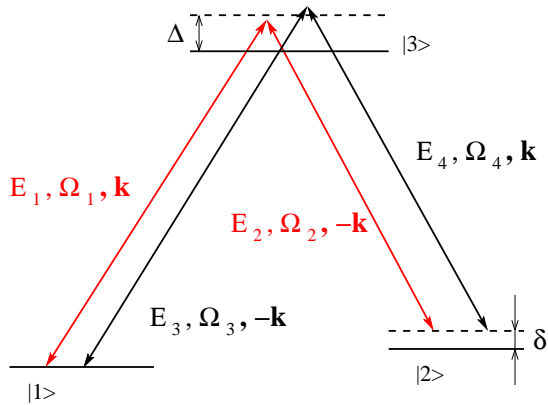}}}
\caption{Schematic representaion of the energy level diagram
and atom - field interaction for the standing wave Raman scheme (SWRS).}
\label{fig1}
\end{figure}

The basic geometry is indicated schematically in Fig. 1. Transitions between
states $\left\vert 1\right\rangle $ and $\left\vert 2\right\rangle $ in the
Raman scheme occur through the common state $\left\vert 3\right\rangle$ using
two field modes. Consider first the effect of fields $E_{1}$ and $E_{2}$.
Field $E_{1}$, having frequency $\Omega_{1}$ and wave vector $\mathbf{k}
_{1}=\mathbf{k=}k\mathbf{\hat{z}}$ drives the $1-3$ transition while field
$E_{2}$, having frequency $\Omega_{2}\approx\Omega_{1}-\omega_{21}$ and wave
vector $\mathbf{k}_{2}\approx-\mathbf{k}$ drives the $2-3$ transition, where
$\omega_{ji}$ is the frequency separation of levels $j$ and $i$ (it is assumed
that $\Omega_{2}/c\approx\Omega_{1}/c$, or, equivalently, that $\omega
_{21}/\omega_{31}\ll1)$. Owing to polarization selection rules or to the fact
that $\omega_{21}$ is greater than the detuning $\Delta=\Omega_{1}-\omega
_{31}$, one can neglect any effects related to field $E_{1}$ driving the $2-3$
transition or field $E_{2}$ driving the $1-3$ transition \cite{single}. If, in
addition, the atom-field detunings on the electronic state transitions are
sufficiently large to enable one to adiabatically eliminate state $\left\vert
3\right\rangle $, one arrives at an effective two-level system in which states
$\left\vert 1\right\rangle $ and $\left\vert 2\right\rangle $ are coupled by a
two-photon "Raman field" having propagation vector $2\mathbf{k}$ and
two-photon detuning $\delta=\Omega_{1}-\Omega_{2}-\omega_{21}$.

Imagine that we start in state $\left\vert 1\right\rangle $. If the initial
state $\left\vert 1\right\rangle $ amplitude is spatially homogeneous, then,
after a two-quantum transition, the final state ( state $\left\vert
2\right\rangle $) amplitude varies as $e^{2ikz}$. Such a state amplitude
amplitude does not correspond to a state $\left\vert 2\right\rangle $
population grating, since the final state density is spatially homogeneous. To
obtain a density grating one can add \emph{another} pair of
counter-propagating fields as shown in Fig. 1. These fields $E_{3}$ and
$E_{4}$ differ in frequency from the initial pair, but the combined two-photon
frequencies are equal,
\begin{equation}
\delta=\Omega_{1}-\Omega_{2}-\omega_{21}=\Omega_{3}-\Omega_{4}-\omega_{21}.
\label{det}
\end{equation}
The propagation vectors are chosen such that $\mathbf{k}_{3}=-\mathbf{k}
_{4}=-\mathbf{k}$. The frequencies of fields $E_{1}$ and $E_{3}$ are taken to
be nearly equal, as are the frequencies of fields $E_{2}$ and $E_{4}$, but it
is assumed that the frequency differences are sufficient to ensure that fields
$E_{1}$ and $E_{3}$ (or $E_{2}$ and $E_{4}$) do not interfere in driving
\emph{single} photon transitions, nor do fields $E_{1}$ and $E_{4}$ (or
$E_{2}$ and $E_{3}$) drive Raman transitions between levels 1 and 2
\cite{cond}. On the other hand, the combined pairs of counter-propagating
fields ($E_{1}$ and $E_{2}$) and ($E_{3}$ and $E_{4}$) \emph{do} interfere in
driving the $1-2$ Raman transition and act as a \textquotedblleft
\emph{standing wave}\textquotedblright\ Raman field which, to lowest order in
the field strengths, leads to a modulation of the final state population given
by $\cos(4kz)$. In this manner, a grating having period $\lambda/4$ is created.

The friction force and diffusion coefficients are calculated using a
semiclassical approach. For $\delta\neq0$, they differ qualitatively from the
corresponding quantities obtained in standard Sisyphus cooling. The physical
origin of the friction force was discussed in I. The calculation can also be
carried out using a quantum Monte-Carlo approach, but the results of such a
calculation are deferred to a future planned publication.

\section{Semi-Classical Equations}

As in I, we consider the somewhat unphysical level scheme in which states
$\left\vert 1\right\rangle $ and $\left\vert 2\right\rangle $ in Fig. 1 have
angular momentum $J=0$, while state $\left\vert 3\right\rangle $ has angular
momentum $J=1$. The field intensities are adjusted such that the Rabi
frequencies $\chi$ (assumed real) associated with all the atom-field
transitions are equal (Rabi frequencies are defined by $-\mu E/2\hbar,$where
$\mu$ is a component of the dipole moment matrix element between ground and
excited states), and the partial decay rate of level 3 to each of levels 1 and
2 is taken equal to $\Gamma/2$ (equal branching ratios for the two
transitions). The fields all are assumed to have the \emph{same} linear
polarization; there is no polarization gradient. The results would be
unchanged if the fields were all $\sigma_{+}$ polarized.

It is assumed that the electronic state detunings are sufficiently large to
satisfy
\begin{equation}
\Omega_{1}-\omega_{31}\approx\Omega_{3}-\omega_{31}\approx\Omega_{2}
-\omega_{32}\approx\Omega_{4}-\omega_{32}\equiv\Delta\gg\Gamma,\chi.
\label{dtineq}
\end{equation}
In this limit and in the rotating-wave approximation, it is possible to
adiabatically eliminate state $\left\vert 3\right\rangle $ and to obtain
equations of motion for ground state density matrix elements. With the same
approximations used in I, one obtains steady-state equations, including
effects related to atomic momentum diffusion resulting from stimulated
emission and absorption, and spontaneous emission. In a field interaction
representation \cite{intrep}, the appropriate equations are \cite{rmn}
\begin{subequations}
\label{dens}
\begin{align}
\alpha\frac{\partial(\rho_{22}-\rho_{11})}{\partial x}  &  =-(\rho_{22}
-\rho_{11})-2i\sigma\cos(x)\,\left[  \rho_{12}-\rho_{21}\right] \, ,
\label{densa}\\
\alpha\frac{\partial\rho_{12}}{\partial x}  &  =-(1+i\mathrm{d})\rho
_{12}-i\sigma\left[  \cos(x)\,(\rho_{22}-\rho_{11})-i\hbar k\sin
(x)\frac{\partial S}{\partial p}\right]  -\cos(x)S/2 \, , \label{densb}\\
\rho_{21}  &  =\rho_{21}^{\ast} \, , \label{densc}
\end{align}
or, in terms of real variables,
\end{subequations}
\begin{subequations}
\label{new}
\begin{equation}
\alpha\frac{\partial}{\partial x}\left(
\begin{array}[c]{c}
u\\
v\\
w
\end{array}
\right)  =\left(
\begin{array}
[c]{lcr}
-1 & \text{\textrm{d}} & 0\\
-\mathrm{d} & -1 & -2\sigma\cos x\\
0 & 2\sigma\cos x & -1
\end{array}
\right)  \left(
\begin{array}[c]{c}
u\\
v\\
w
\end{array}
\right)  -\left(
\begin{array}[c]{c}
\cos x\,S+2\hbar k\sigma\sin x\frac{\partial S}{\partial p}\\
0\\
0
\end{array}
\right)  , \label{new-b}
\end{equation}
where the total population $S$ evolves as
\end{subequations}
\begin{equation}
\frac{\partial S}{\partial t}=\frac{7}{5}\hbar^{2}k^{2}\Gamma^{\prime}
\frac{\partial^{2}S}{\partial p^{2}}-4\Gamma^{\prime}\sigma\hbar k\sin
x\frac{\partial u}{\partial p}\,-\frac{3}{5}\hbar^{2}k^{2}\Gamma^{\prime}\cos
x\frac{\partial^{2}u}{\partial p^{2}}\,, \label{newfp}
\end{equation}
and
\begin{subequations}
\label{def}
\begin{align}
u  &  =\rho_{12}+\rho_{21} \, ,\label{defa}\\
v  &  =i\left(  \rho_{21}-\rho_{12}\right) \, ,\label{defb}\\
w  &  =\rho_{22}-\rho_{11} \, ,\label{defc}\\
S  &  =\rho_{11}+\rho_{22} \, , \label{defd}
\end{align}
with
\end{subequations}
\begin{subequations}\label{par}
\begin{align}
x  &  =2kz,\label{para}\\
\mathrm{d}  &  =\frac{\delta}{2\Gamma^{\prime}}\label{parb}\\
\alpha &  =k\mathrm{v}/\Gamma^{\prime},\label{parc}\\
\sigma &  =\Delta/\Gamma,\label{pard}\\
\Gamma^{\prime}  &  =\Gamma\chi^{2}/\left[  \Delta^{2}+\left(  \Gamma
/2\right)  ^{2}\right]  \sim\chi^{2}\Gamma/\Delta^{2}. \label{pare}
\end{align}
Each of the functions $u,v,w,S$ are now functions of the $z$-component of
momentum $p=M\mathrm{v}$ ($M$ is the atom's mass and $\mathrm{v}$ is the
$z$-component of atomic velocity) as well as $x$, but it is assumed in this
semiclassical approach that $S$ is position independent. The parameter
$\sigma=\Delta/\Gamma$ is assumed to be large compared with unity.

It will also prove useful to define dimensionless frequencies normalized to
$\omega_{r}$, momenta normalized to $\hbar k$, and energies normalized to
$\hbar\omega_{r}$, where $\omega_{r}$ is the recoil frequency
\end{subequations}
\begin{equation}
\omega_{r}=\hbar k^{2}/2M, \label{omrec}
\end{equation}
such that $\tilde{\delta}=\delta/\omega_{r}$, $\tilde{\Gamma}=\Gamma
/\omega_{r}$, $\tilde{\Gamma}^{\prime}=\Gamma^{\prime}/\omega_{r}$,
$\tilde{\chi}_{eff}=\chi_{eff}/\omega_{r}$ \, \{$\chi_{eff}=\chi^{2}
\Delta/\left[  \Delta^{2}+\left(  \Gamma/2\right)  ^{2}\right]  =\Gamma
^{\prime}\sigma)$ is an effective two-photon Rabi frequency\}, $\bar
{p}=p/\hbar k$, etc. In terms of these quantities,
\begin{subequations}
\label{nor}
\begin{align}
\tilde{\chi}_{eff}  &  =\frac{\chi^{2}\Delta/\omega_{r}}{\Delta^{2}+\left(
\Gamma/2\right)  ^{2}}\equiv I \, ,\label{nora}\\
\tilde{\Gamma}^{\prime}  &  =I/\sigma \, ,\label{norb}\\
\mathrm{d}  &  =\tilde{\delta}\sigma/2I \, ,\label{norc}\\
\alpha &  =\bar{p}/\tilde{\Gamma}^{\prime}=2\sigma\bar{p}/I \, . \label{nord}
\end{align}
Note that $I$ is the effective coupling strength normalized to the recoil
frequency.

Equation (\ref{new-b}) can be written in matrix form as
\end{subequations}
\begin{equation}
\alpha\frac{d\mathbf{B}(x)}{dx}=-\left[  \mathbf{A}_{1}+2\sigma\cos
x\,\mathbf{A}_{2}\right]  \mathbf{B}(x)-\mathbf{\Lambda}(x), \label{Matrix1}
\end{equation}
where
\begin{equation}
\mathbf{B}(x)=\left(
\begin{array}[c]{c}
u\\
v\\
w
\end{array}
\right)  \,,\,\,\mathbf{A}_{1}=\left(
\begin{array}[c]{lcr}
1 & -\mathrm{d} & 0\\
\mathrm{d} & 1 & 0\\
0 & 0 & 1
\end{array}
\right)  \,,\,\,\mathbf{A}_{2}=\left(
\begin{array}[c]{lcr}
0 & 0 & 0\\
0 & 0 & 1\\
0 & -1 & 0
\end{array}
\right) \, , \label{mat}
\end{equation}
\begin{equation}
\mathbf{\Lambda}(x)=\left[  \cos x\,S(p)+2\sigma\hbar k\sin x\frac{\partial
S}{\partial p}\right]  \left(
\begin{array}[c]{l}
1\\
0\\
0
\end{array}
\right)  \,. \label{vect}
\end{equation}

It should be noted that Eq. (\ref{Matrix1}) differs \textit{qualitatively}
from the corresponding equation encountered in high intensity laser theory.
Owing to the fact that decay of $u,v,w$ is linked to spontaneous emission, the
decay parameters depend on field intensity. When all frequencies are
normalized to the optical pumping rate $\Gamma^{\prime}$, the effective
coupling strength $\sigma$ is actually independent of field strength;
moreover, since it is assumed that $\sigma>1$, one is always in a "high
intensity" limit. In contrast to the equations describing conventional
Sisyphus cooling or high intensity laser theory, there is a source term for
$u$, but no source term for the population difference $w$.

The formal solution of Eq. (\ref{Matrix1}) satisfying boundary conditions
resulting in a periodic solution is
\begin{equation}
\mathbf{B}(x)=-\frac{1}{\alpha}\int_{-\infty}^{x}dx^{\prime}e^{-\mathbf{A}
_{1}(x-x^{\prime})/\alpha}\left[  2\sigma\cos x^{\prime}\mathbf{A}
_{2}\mathbf{B}(x^{\prime})+\mathbf{\Lambda}(x^{\prime})\right]  ,
\label{int-form1}
\end{equation}
which, in terms of components, can be written as
\begin{subequations}
\label{int}
\begin{align}
u(x)  &  =-\int_{0}^{\infty}d\tau e^{-\tau}\left\{  2\sigma\cos(\alpha
\tau-x)\sin\left[  \mathrm{d}\tau\right]  w(x-\alpha\tau)+s(x-\alpha\tau
)\cos\left[  \mathrm{d}\tau\right]  \right\} \, ,\label{inta}\\
v(x)  &  =-\int_{0}^{\infty}d\tau e^{-\tau}\left\{  2\sigma\cos(\alpha
\tau-x)\cos\left[  \mathrm{d}\tau\right]  w(x-\alpha\tau)-s(x-\alpha\tau
)\sin\left[  \mathrm{d}\tau\right]  \right\} \, ,\label{intb}\\
w(x)  &  =2\sigma\int_{0}^{\infty}d\tau e^{-\tau}\cos(\alpha\tau
-x)v(x-\alpha\tau), \label{intc}
\end{align}
where
\end{subequations}
\begin{equation}
s(x)=\cos x\,S(p)+2\sigma\hbar k\sin x\frac{\partial S}{\partial p}\,.
\label{ksi}
\end{equation}
Substituting $v(x)$ into the equation for $w(x)$ we obtain
\begin{align}
w(x)  &  =-4\sigma^{2}\int_{0}^{\infty}d\tau e^{-\tau}\cos(\alpha\tau
-x)\int_{0}^{\infty}d\tau^{\prime}e^{-\tau^{\prime}}\cos(\alpha(\tau
+\tau^{\prime})-x)\cos\left[  \mathrm{d}\tau^{\prime}\right]  w\left[
x-\alpha(\tau+\tau^{\prime})\right] \nonumber\\
&  +2\sigma\int_{0}^{\infty}d\tau e^{-\tau}\cos(\alpha\tau-x)\int_{0}^{\infty
}d\tau^{\prime}e^{-\tau^{\prime}}s\left[  x-\alpha(\tau+\tau^{\prime})\right]
\sin\left[  \mathrm{d}\tau^{\prime}\right]  . \label{weq1}
\end{align}

Once the solution for $w(x)$ is obtained, it is substituted into Eq.
(\ref{inta}) to determine $u(x)$ and the solution for $u(x)$ substituted into
Eq.\ (\ref{newfp}) for $\partial S/\partial t.$ The resultant equation is
averaged over a wavelength resulting in
\begin{equation}
\frac{\partial S}{\partial t}=\frac{7}{5}\hbar^{2}k^{2}\Gamma^{\prime}
\frac{\partial^{2}S}{\partial p^{2}}-4\Gamma^{\prime}\sigma\hbar
k\frac{\partial}{\partial p}\eta_{1}-\frac{3}{5}\hbar^{2}k^{2}\Gamma^{\prime
}\frac{\partial^{2}}{\partial p^{2}}\eta_{2}\,, \label{fp1}
\end{equation}
where
\begin{subequations}
\label{eta}
\begin{align}
\eta_{1}  &  =\overline{u(x)\sin x}=\frac{1}{2\pi}\int_{0}^{2\pi}dxu(x)\sin
x\,,\label{eta1}\\
\eta_{2}  &  =\overline{u(x)\cos x}=\frac{1}{2\pi}\int_{0}^{2\pi}dxu(x)\cos
x\,, \label{eta2}
\end{align}
and the bar indicates a spatial average ($\bar{S}=S$, by assumption$)$.
Equation (\ref{fp1}) is then compared with the FokkerPlanck equation
\end{subequations}
\begin{equation}
\frac{\partial S}{\partial t}=\frac{\partial}{\partial p}\left[  -\bar
{F}\,S+\bar{D}_{st}\frac{\partial S}{\partial p}+\frac{\partial}{\partial
p}\left[  \bar{D}_{sp}S\right]  \right]  , \label{fp3}
\end{equation}
to extract the spatially averaged friction $\bar{F},$ stimulated diffusion
$\bar{D}_{st}$, and spontaneous diffusion $\bar{D}_{sp}$ coefficients.

\section{Solutions}

\subsection{Backward recursion method}

Equation (\ref{weq1}) can be solved using Fourier series and a backwards
recursion scheme \cite{thaler,zieg}, as outlined in the Appendix. In this
manner one obtains
\begin{subequations}
\label{avg}
\begin{align}
\bar{F}  &  =-2\hbar k\sigma\Gamma^{\prime}\xi_{f} \, ,\label{avga}\\
\bar{D}_{st}  &  =4\hbar^{2}k^{2}\sigma^{2}\Gamma^{\prime}\xi_{st} \, ,
\label{avgb}\\
\bar{D}_{sp}  &  =\hbar^{2}k^{2}\Gamma^{\prime}\left(  \frac{7}{5}+\frac
{3}{10}\xi_{sp}\right) \, , \label{avgc}
\end{align}
where $\xi_{f},$ $\xi_{sp}$, $\xi_{st}$ are given as Eq. (A17) in the
Appendix. An analytic solution for the $\xi$s can be found only for
$\mathrm{d\ }=0$\, $\left[  \xi_{f}(\mathrm{d\ }=0)=\alpha
(1+\alpha^{2})^{-1} \right.$,
$\left. \text{ }\xi_{sp}(0)=\xi_{st}(0)=(1+\alpha^{2})^{-1} \right]$;
otherwise,
these quantities must be obtained via the recursive solutions. The effective
field strength parameter in this problem is $\sigma$ and one might expect that
$\sigma/\alpha$ recursions are needed to solve Eqs. (\ref{mrec}) accurately
\cite{thaler,zieg}. Actually, the number of recursions required depends in a
somewhat complicated manner on the values of several parameters. Each
recursion introduces resonances at specific values of $\alpha$ which can be
interpreted as Stark-shifted, velocity tuned resonances \cite{zieg2}. For
example, with $\mathrm{d\ }=0,$ the lowest order recursive solution has a very
strong (proportional to $\sigma^{2}$), narrow resonance at $\alpha^{2}=3/5$,
but the second order approximation removes the divergence, while introducing
yet a second resonance. Some of these velocity tuned resonances are seen in
some of the graphs presented below. For $\mathrm{d}>1$, an upper bound for the
number of recursions required to map out all the resonances is of order
$\left(  \sigma/\mathrm{d}\right)  \left(  \sigma/\mathrm{\alpha}\right)  ;$
for $\mathrm{d}\ll1$ or $\mathrm{d}/\sigma\gg1$ only a few terms are needed.
Even if only a single recursion is needed, the general expressions for
$\xi_{f},$ $\xi_{sp}$, $\xi_{st}$ are still fairly complicated.

For $\mathrm{d}\ll1$ one finds corrections of order $\mathrm{d}^{2}$ to the
analytical results, but owing to their complexity, these expressions are not
given here. For $\mathrm{d}/\sigma\gg1,$ one finds that, near $\alpha=0$
\end{subequations}
\begin{subequations}
\label{cs}
\begin{align}
\xi_{f}  &  \sim\frac{2\alpha}{1+4\alpha^{2}}\left(  \frac{\sigma}{\mathrm{d}
}\right)  ^{2} \, ,\label{csa}\\
\xi_{st}  &  \sim\frac{1}{1+4\alpha^{2}}\left(  \frac{\sigma}{\mathrm{d}
}\right)  ^{2} \, ,\label{csb}\\
\xi_{sp}  &  \sim\frac{\left(  3+8\alpha^{2}\right)  }{1+4\alpha^{2}}\left(
\frac{\sigma}{\mathrm{d}}\right)^{2} \, ,\label{csc}
\end{align}
and near $\alpha=\pm\mathrm{d}$
\end{subequations}
\begin{subequations}
\label{cd}
\begin{align}
\xi_{f}  &  \sim\frac{f_{\pm}}{2\left(  1+f_{\pm}^{2}+\sigma^{2}\right)
} \, ,\label{cda}\\
\xi_{st}  &  \sim\frac{1}{2\left(  1+f_{\pm}^{2}+\sigma^{2}\right)
} \, ,\label{cdb}\\
\xi_{sp}  &  \sim\frac{1+\sigma^{2}}{2\left(  1+f_{\pm}^{2}+\sigma^{2}\right)
} \, ,\label{cdc}
\end{align}
where $f_{\pm}=\left(\alpha\mp\mathrm{d}\right)$.
\begin{figure}[th]
\centerline{\scalebox{0.4}{\includegraphics{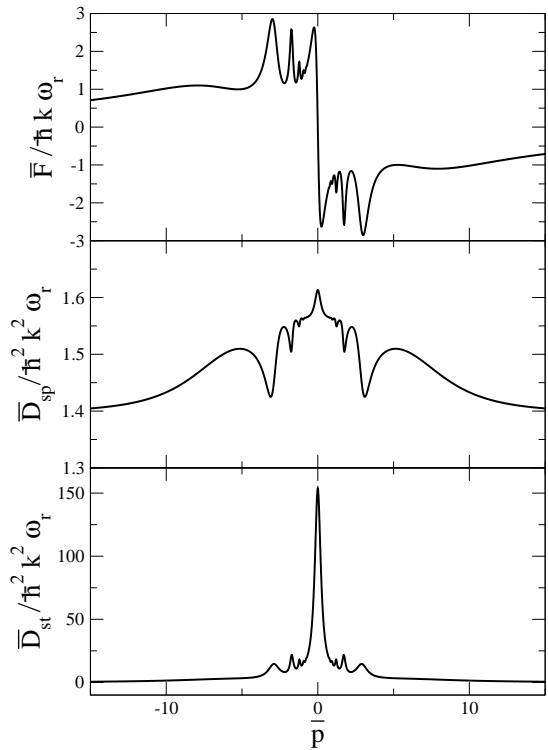}}}
\caption{Averaged force and diffusion coefficients in dimensionless units as a
function of scaled momentum $\bar{p}=p/\hbar k$ for $I=10$, $\sigma=10$, and
$\tilde{\delta}=20$.}
\end{figure}

In the limit $\mathrm{d}\gg\sigma>1$, the friction force as a function of
$\alpha$ contains three dispersion-like structures centered at $\alpha
=0,\pm\mathrm{d}$. This implies that atoms can be cooled to these values of
$\alpha$. The amplitude of the component centered at $\alpha=0$ is of order
$\sigma^{2}/\mathrm{d}^{2}$ while its width is of order unity. On the other
hand, the amplitude of the components centered at $\alpha=\pm\mathrm{d}$ are
of order $1/\sigma$ while their width are of order $\sigma$. It is shown below
that the central peak in the momentum distribution is negligible compared with
the two side peaks in the limit $\mathrm{d}\gg\sigma>1$; that is, in this
limit cooling occurs more efficiently to velocities $\alpha=\pm\mathrm{d}$ for
which the atoms are Doppler shifted into resonance with the two-photon
transitions connecting levels 1 and 2. The width of the $\alpha=\pm\mathrm{d}$
components is similar to that found in sub-Doppler cooling in magnetic fields
(MILC) \cite{sismag}; in both MILC and the SWRS, the qualitative dependence
for the friction coefficient as a function of $\alpha$ is similar to that
found in sub-Doppler cooling using "corkscrew" polarization \cite{cohen}. As
such, the curve is "power broadened," since the effective field strength in
the problem is $\sigma.$

It is tempting to consider the contribution to the friction force near
$\alpha=\mathrm{d}$ as arising from the \textit{single} pair of fields
$\left(  E_{1}\text{ and }E_{2}\right)  $, since these fields are nearly
resonant with the 1-2 transition in a reference frame moving at
$2k\mathrm{v=\delta}$. Tempting as it may be, this interpretation is wrong,
since we have already shown in I that, for a single pair of Raman fields, the
friction force vanishes \textit{identically}, regardless of detuning. Thus, it
is necessary that the second pair of fields be present, even if they are far
off resonance with atoms satisfying $2k\mathrm{v=\delta}$. The main effect of
the second pair of fields is to cancel the contribution to the force from the
population difference between levels 1 and 2 (see Appendix A in I), leaving
the contribution from the 1-2 coherence only ($u=\rho_{12}+\rho_{21}$). Near
$2k\mathrm{v=\delta}$ the major contribution to $u$ \textit{does} come from
atoms that are nearly resonant with the 1-2 transition in a reference frame
moving at $2k\mathrm{v=\delta,}$ but the scattering of the second pair of
fields $\left(  E_{3}\text{ and }E_{4}\right)  $ from the population
difference created by the resonant pair of fields modifies the net force on
the atoms. In some sense, one can view the second pair of fields as enabling
the cooling at $2k\mathrm{v=\delta}$. Note that the magnitude of the damping
coefficient is down by $\sigma^{2}$ from that at $\mathrm{d=0}$; it is of the
same order as that found in sub-Doppler cooling using "corkscrew" polarization
\cite{cohen}.
\begin{figure}[t]
\centerline{\scalebox{0.4}{\includegraphics{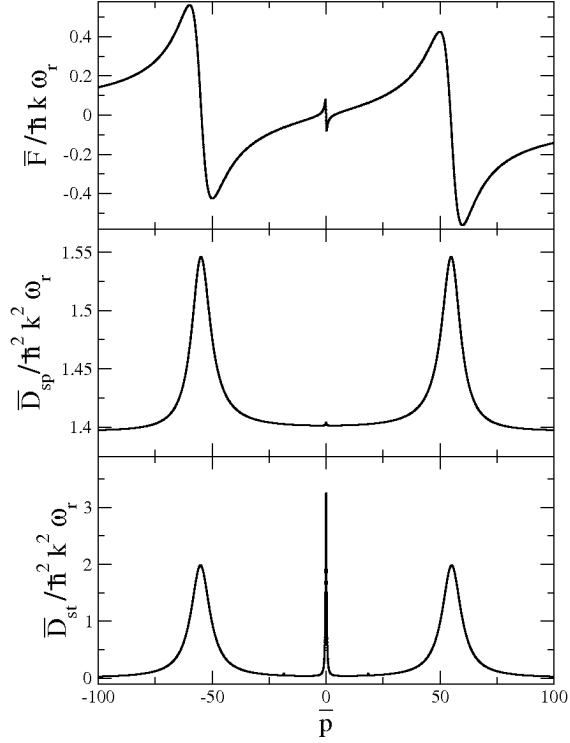}}}
\caption{Same as Fig. 2, with $I=10$, $\sigma=10$, and
$\tilde{\delta}=220$.}
\end{figure}

For arbitrary values of $\alpha$ and $\delta$, with $\sigma$ of order 10, the
recursive solution converges very rapidly for most values of $\alpha$ and
numerical solutions can be obtained quickly and easily. Two examples are shown
in Figs. 2 and 3, where the averaged friction force $\bar{F}$ in units of
$\hbar k\omega_{r}$ and the averaged diffusion coefficients $\bar{D}_{st}$ and
$\bar{D}_{sp}$ in units $\hbar^{2}k^{2}\omega_{r}$ are plotted as a function
of the scaled momentum $\bar{p}=p/\hbar k$. In terms of the $\xi$s defined by
Eqs (\ref{avg}), these quantities can be written as
\end{subequations}
\begin{align*}
\bar{F}/\hbar k\omega_{r}  &  =-2I\xi_{f}(\alpha=2\sigma\bar{p}/I) \, ,\\
\bar{D}_{st}/\hbar^{2}k^{2}\omega_{r}  &  =4\sigma I\xi_{st}(\alpha
=2\sigma\bar{p}/I) \, ,\\
\bar{D}_{sp}/\hbar^{2}k^{2}\omega_{r}  &  =I\left[  \frac{7}{5}+\frac{3}
{10}\xi_{sp}(\alpha=2\sigma\bar{p}/I)\right]  /\sigma.
\end{align*}
In Fig. 2, $I=10$, $\sigma=10$, and $\tilde{\delta}=\delta/\omega_{r}=20$. One
sees in these curves a number of velocity tuned resonances under a
"power-broadened" envelope \cite{zieg2}. In Fig. 3, $I=10$, $\sigma=10$, and
$\tilde{\delta}=220$, implying that $\mathrm{d=110}$ and $\sigma
/\mathrm{d=1/11}$. In this limit Eqs. (\ref{cs}), (\ref{cd}) are valid and we
see three contributions to the averaged force and diffusion coefficients. The
values of the force and diffusion coefficients near the Doppler tuned
resonances at $\bar{p}=\pm\tilde{\delta}/4$ are typical of corkscrew
polarization cooling \cite{cohen}, and the ratio of the force to diffusion
coefficient is of order $1/\hbar k$. On the other hand, this ratio is of order
$1/\hbar k\sigma$ near $\bar{p}=0$, a result that is typical of Sisyphus
cooling in a lin$\perp$lin geometry; however, both the friction and diffusion
coefficients are smaller than those in conventional Sisyphus cooling by a
factor $\left(  \sigma/\mathrm{d}\right)  ^{2}$ when $\sigma/\mathrm{d\ll1.}$
As a consequence, the cooling is dominated by the contributions near $\bar
{p}=\pm\tilde{\delta}/4$ when $\sigma/\mathrm{d\ll1}$.

\subsection{Iterative Solution}

Since the effective field strength is always greater than unity, perturbative
solutions of Eqs (\ref{Matrix1}) are not of much use. However, one can get a
very rough qualitative estimate of the dependence on detuning of the friction
and diffusion coefficients near $\alpha=0$ by considering an iterative
solution of Eqs. (\ref{Matrix1}) in powers of $\alpha$. This will work only in
the limit that $\alpha<1,$ so it cannot correctly reproduce the contributions
to the friction and diffusion coefficients at $\alpha=\pm\mathrm{d}$ when
$\mathrm{d}\gg\sigma>1$. The iterative solution is useful mainly when
$\mathrm{d}^{2}\lesssim\sigma^{2}$, since, in this limit, the dominant
contribution to the momentum distribution comes from the region near
$\alpha=0$.

The iterative solution is straightforward, but algebraically ugly. To order
$\alpha$, one obtains from Eqs. (\ref{Matrix1})
\begin{equation}
\mathbf{B}(x)=-\mathbf{A}^{-1}(x)\left\{  \left[  \mathbf{\Lambda}(x)\right]
-\alpha\frac{d}{dx}\left(  \mathbf{A}^{-1}(x)\left[  \mathbf{\Lambda
}(x\right]  \right)  \right\}  , \label{bxp}
\end{equation}
where $\mathbf{A}(x)=\left[  \mathbf{A}_{1}+2\sigma\cos x\,\mathbf{A}
_{2}\right]  $. When the $u$ component of $\mathbf{B}(x)$ is extracted from
this solution and the result is substituted into Eqs. (\ref{eta}), all the
integrals can be carried out analytically and one finds
\begin{align*}
\eta_{1}  &  =-2\xi_{1}^{\left(  0\right)  }g-2\alpha\xi_{1}^{\left(
1\right)  }S \, ,\\
\eta_{2}  &  =-2\xi_{2}^{\left(  0\right)  }S+2\alpha\xi_{2}^{\left(
1\right)  }g \, ,
\end{align*}
where
\begin{subequations}
\label{ps}
\begin{align}
g  &  =2\sigma\hbar k\frac{\partial S}{\partial p} \, ,\label{psa}\\
\xi_{1}^{\left(  0\right)  }  &  =1-\frac{2\mathrm{d}^{2}}{\gamma_{d}\left(
\gamma_{d}+\gamma_{\sigma}\right)  } \, ,\label{psb}\\
\xi_{2}^{\left(  0\right)  }  &  =1-\frac{2\mathrm{d}^{2}}{\gamma_{\sigma
}\left(  \gamma_{d}+\gamma_{\sigma}\right)  } \, ,\label{psc}\\
\xi_{1}^{\left(  1\right)  }  &  =\xi_{2}^{\left(  1\right)  }=1-\frac
{\mathrm{d}^{2}\left[  \gamma_{\sigma}^{2}\gamma_{d}^{2}+\gamma_{d}^{2}
+\gamma_{\sigma}^{2}\right]  }{\gamma_{\sigma}^{3}\gamma_{d}^{3}}
\, ,\label{psd}\\
\gamma_{d}  &  =\sqrt{1+\mathrm{d}^{2}}\text{\, , \ \ \ \ }\gamma_{\sigma}
=\sqrt{1+\mathrm{d}^{2}+4\sigma^{2}}\text{.} \label{pse}
\end{align}
By comparing Eqs. (\ref{fp1}), (\ref{fp3}) and neglecting the contribution from
the second term in the equation for $\eta_{2}$ (since it is of relative order
$\hslash k/p),$ we extract the spatially averaged friction and diffusion
coefficients
\end{subequations}
\begin{subequations}
\label{pert}
\begin{align}
\bar{F}  &  =-2\hbar k\sigma\Gamma^{\prime}\alpha\xi_{1}^{\left(  1\right)
} \, ,\label{perta}\\
\bar{D}_{st}  &  =4\hbar^{2}k^{2}\sigma^{2}\Gamma^{\prime}\xi_{1}^{\left(
0\right)  } \, ,\label{pertb}\\
\bar{D}_{sp}  &  =\hbar^{2}k^{2}\Gamma^{\prime}\left(  \frac{7}{5}+\frac
{3}{10}\xi_{2}^{\left(  0\right)  }\right) \, .\label{pertc}
\end{align}
These are all even functions of the detuning $\mathrm{d}$.

The spatially averaged form factors $\xi_{1}^{\left(  0\right)  },\xi
_{2}^{\left(  0\right)  },\xi_{1}^{\left(  1\right)  }$ are equal to unity for
$\mathrm{d}=0$, but vary as
\end{subequations}
\begin{equation}
\xi_{2}^{\left(  0\right)  }\approx\xi_{1}^{\left(  0\right)  }\approx
(3/2)\xi_{1}^{\left(  1\right)  }\sim3\sigma^{2}/\mathrm{d}^{2} \label{zetasy}
\end{equation}
for $\mathrm{d}\gg\mathrm{d}/\sigma\gg1$, in agreement with Eqs. (\ref{cs}).
In this limit, both $\bar{F}$ and $\bar{D}_{ind}$ approach zero, but $\bar
{D}_{sp}$ approaches a finite value since Rayleigh scattering of the fields is
independent of $\delta$ for $\delta\ll\Delta$. The friction force when
$\mathrm{d}\gg\mathrm{d}/\sigma\gg1$ is given by
\[
\bar{F}\sim-4\Gamma^{\prime}\hbar k\sigma\alpha\sigma^{2}/\mathrm{d}
^{2}=-16\hbar k^{2}v\left[  \Gamma\left(  \Delta\delta/\Gamma^{2}\right)
\right]  \left(  \chi^{2}/\Delta\right)  ^{2}/\delta^{3}.
\]
This equation is strangely reminiscent of the equation for Doppler cooling of
two-level atoms by an off-resonant standing wave field where one finds
\[
\bar{F}_{DC}\approx-4\hbar k\left(  kv\right)  \Gamma\left(  \chi^{2}\right)
/\Delta^{3},
\]
taking into account the fact that twice the momentum is transferred in a
two-photon process. For the expressions to agree, one must associate a
"two-photon spontaneous scattering rate" $\Gamma_{tp}=\Gamma\left(
\Delta\delta/\Gamma^{2}\right)  $ with the Raman transitions.

Of course, if $\mathrm{d}/\sigma\gg1,$ the contributions to the friction and
diffusion coefficients near $\alpha=\pm\mathrm{d}$ become dominant insofar as
they affect the momentum distribution. In this limit one cannot use the
iterative solution to estimate the equilibrium temperature since the
contributions from higher velocity components play a significant role.

\subsection{Dressed State Solution}

The effective Hamiltonian for the SWRS, neglecting decay is
\begin{equation}
H=\frac{\hbar}{2}\left(
\begin{array}
[c]{cc}
\delta & 4\chi_{eff}\cos x\\
4\chi_{eff}\cos x & -\delta
\end{array}
\right)  . \label{hf}
\end{equation}
By diagonalizing this Hamiltonian one obtains semiclassical dressed states
whose energies, as a function of $x$ are said to characterize the optical
potentials associated with this problem. It turns out that the use of dressed
states in the SWRS is of somewhat limited use, owing to \textit{nonadiabatic}
coupling between the potentials. Nevertheless, the dressed states do
provide additional insight to the cooling dynamics.
\begin{figure}[th]
\centerline{\scalebox{0.5}{\includegraphics{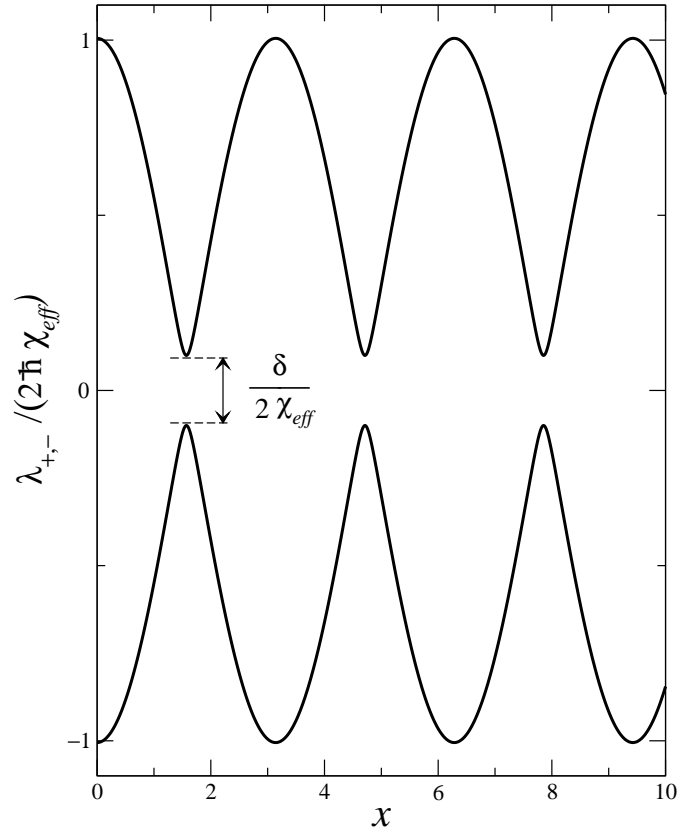}}}
\caption{Dressed state potentials.} \label{optpot}
\end{figure}

The eigenvalues of $H$ are given by
\begin{equation}
\lambda_{\pm}=\pm\hbar R/2, \text{ \ \ \ }R=\sqrt{\delta^{2}+R_{0}^{2}},
\label{eigen}
\end{equation}
along with eigenkets
\begin{subequations}
\label{ket}
\begin{align}
\left\vert A\right\rangle  &  =c\left\vert 1\right\rangle +s\left\vert
2\right\rangle, \label{keta}\\
\left\vert B\right\rangle  &  =-s\left\vert 1\right\rangle +c\left\vert
2\right\rangle , \label{ketb}
\end{align}
where
\end{subequations}
\begin{subequations}
\label{dp}
\begin{align}
c  &  =\cos\theta, \text{ \ \ }s=\sin\theta, \label{dpa}\\
R_{0}  &  =4\chi_{eff}\cos x, \label{dpb}\\
\cos(2\theta)  &  =\delta/R, \text{ \ \ \ \ }\sin(2\theta)=R_{0}/R, \text{
\ \ }0\leq\theta\leq\pi/2. \label{dpc}
\end{align}
The optical potentials are sketched in Fig. 4. As $\delta/2\chi_{eff}$ tends
towards zero, the potentials "touch" whenever $\cos x=0$. As is seen below,
nonadiabatic transitions occur at such points~\cite{moddress}.

Defining dressed state amplitudes via
\end{subequations}
\begin{equation}
\mathbf{a}_{D}=\mathbf{Ta} \label{dsa}
\end{equation}
with
\[
\mathbf{a}_{D}=\left(
\begin{array}[c]{c}
a_{A}\\
a_{B}
\end{array}
\right)  ;\text{ \ \ }\mathbf{a}_{D}=\left(
\begin{array}[c]{c}
a_{1}\\
a_{2}
\end{array}
\right)  ;\text{ \ \ \ \ }\mathbf{T=}\left(
\begin{array}[c]{cc}
c & s\\
-s & c
\end{array}
\right)
\]
and a dressed state density matrix $\mathbf{\rho}_{D}=\mathbf{a}_{D}
\mathbf{a}_{D}^{\dag}$, one can transform Eqs. (\ref{new}), (\ref{def})
into the dressed basis as
\begin{subequations}
\label{drs}
\begin{align}
\alpha\frac{\partial w_{D}}{\partial x}  &  =-w_{D}+\sin(2\theta)\left[
\cos(x)S+2\sigma\hbar k\sin\left(  x\right)  \frac{\partial S}{\partial
p}\right]\nonumber\\
& -\frac{\sin(2\theta)v_{D}}{2\sigma}+2\alpha\frac{\partial\theta
}{\partial x}\left(  \rho_{AB}+\rho_{BA}\right), \label{drsa}\\
\alpha\frac{\partial\rho_{AB}}{\partial x}  &  =-\left(  1+iD\right)
\rho_{AB}-\cos(2\theta)\left[  \frac{1}{2}\cos(x)S+\sigma\hbar k\sin\left(
x\right)  \frac{\partial S}{\partial p}\right] \nonumber\\
&  -\frac{i\sin(2\theta)w_{D}}{2\sigma}+\alpha\frac{\partial\theta}{\partial
x}\left(  \rho_{BB}-\rho_{BA}\right), \label{drsc}\\
\frac{\partial S}{\partial t}  &  =\frac{7}{5}\hbar^{2}k^{2}\Gamma^{\prime
}\frac{\partial^{2}S}{\partial p^{2}}-4\Gamma^{\prime}\sigma\hbar k\sin
x\frac{\partial\left[  \cos(2\theta)u_{D}-\sin(2\theta)w_{D}\right]
}{\partial p}\nonumber\\
&  \,-\frac{3}{5}\hbar^{2}k^{2}\Gamma^{\prime}\cos x\frac{\partial^{2}\left[
\cos(2\theta)u_{D}-\sin(2\theta)w_{D}\right]  }{\partial p^{2}}, \label{drse}\\
u_{D}  &  =\rho_{AB}+\rho_{BA}, \text{ \ \ }v_{D}=i\left(  \rho_{BA}-\rho
_{AB}\right), \text{ \ \ }w_{D}=\rho_{BB}-\rho_{AA}, \label{drsf}\\
\rho_{BA}  &  =\rho_{AB}^{\ast}, \label{drsg}\\
D  &  =R/2\Gamma^{\prime} .\label{drsh}
\end{align}
For $\sigma\gg1$, the terms varying as $\sigma^{-1}$ can be dropped. If one
also neglects the nonadiabatic coupling proportional to $\partial
\theta/\partial x,$ Eqs. (\ref{drs}) have the \textit{remarkable} property
that, even in the presence of dissipation, the equations for the dressed state
coherences and populations are completely decoupled. Assuming for the moment
that such an approximation is valid, one has the immediate solution
\end{subequations}
\begin{subequations}
\label{dss}
\begin{align}
\rho_{AB}  &  =-\left(  \alpha\right)  ^{-1}\int_{-\infty}^{x}dx^{\prime}
\cos\left[  2\theta\left(  x^{\prime}\right)  \right]  \left[  \frac{1}{2}
\cos(x^{\prime})S+\sigma\hbar k\sin\left(  x^{\prime}\right)  \frac{\partial
S}{\partial p}\right]  \exp\left[  -\left(  1+iD\right)  \left(  x-x^{\prime
}\right)  /\alpha\right]  ;\label{dssa}\\
w_{AB}  &  =\left(  \alpha\right)  ^{-1}\int_{-\infty}^{x}dx^{\prime}
\sin\left[  2\theta\left(  x^{\prime}\right)  \right]  \left[  \cos(x^{\prime
})S+2\sigma\hbar k\sin\left(  x^{\prime}\right)  \frac{\partial S}{\partial
p}\right]  \exp\left[  -\left(  x-x^{\prime}\right)  /\alpha\right]  .
\label{dssb}
\end{align}
It then follows from Eqs. (\ref{drs}), and (\ref{fp3}) that the spatially
averaged friction and diffusion coefficients are given by
\end{subequations}
\begin{subequations}
\label{dcf}
\begin{align}
\bar{F}  &  =4\hbar k\sigma\Gamma^{\prime}\overline{\sin\left(  x\right)
\left\{  \cos\left[  2\theta\left(  x\right)  \right]  \left[  C(x)+C^{\ast
}(x)\right]  -\sin\left[  2\theta\left(  x\right)  \right]  A(x)\right\}
}, \label{dcfa}\\
\bar{D}_{st}  &  =-8\hbar^{2}k^{2}\sigma^{2}\Gamma^{\prime}\overline
{\sin\left(  x\right)  \left\{  \cos\left[  2\theta\left(  x\right)  \right]
\left[  D(x)+D^{\ast}(x)\right]  -\sin\left[  2\theta\left(  x\right)
\right]  B(x)\right\}  }, \label{dcfb}\\
\bar{D}_{sp}  &  =\hbar^{2}k^{2}\Gamma^{\prime}\left(  \frac{7}{5}+\frac
{3}{10}\overline{\cos\left(  x\right)  \left\{  \cos\left[  2\theta\left(
x\right)  \right]  \left[  C(x)+C^{\ast}(x)\right]  -\sin\left[
2\theta\left(  x\right)  \right]  A(x)\right\}  }\right), \label{dcfc}
\end{align}
where
\end{subequations}
\begin{align*}
A(x)  &  =\left(  \alpha\right)  ^{-1}\int_{-\infty}^{x}dx^{\prime}\sin\left[
2\theta\left(  x^{\prime}\right)  \right]  \cos(x^{\prime})\exp\left[
-\left(  x-x^{\prime}\right)  /\alpha\right]  ,\\
B(x)  &  =\left(  \alpha\right)  ^{-1}\int_{-\infty}^{x}dx^{\prime}\sin\left[
2\theta\left(  x^{\prime}\right)  \right]  \sin(x^{\prime})\exp\left[
-\left(  x-x^{\prime}\right)  /\alpha\right]  ,\\
C(x)  &  =-\left(  2\alpha\right)  ^{-1}\int_{-\infty}^{x}dx^{\prime}
\cos\left[  2\theta\left(  x^{\prime}\right)  \right]  \cos(x^{\prime}
)\exp\left[  -\left(  1+iD\right)  \left(  x-x^{\prime}\right)  /\alpha
\right]  ,\\
D(x)  &  =-\left(  2\alpha\right)  ^{-1}\int_{-\infty}^{x}dx^{\prime}
\cos\left[  2\theta\left(  x^{\prime}\right)  \right]  \sin(x^{\prime}
)\exp\left[  -\left(  1+iD\right)  \left(  x-x^{\prime}\right)  /\alpha
\right]  ,
\end{align*}
and the bar indicates a spatial average. In general, the integrals and spatial
averages must be calculated numerically.

In contrast to other dressed state theories, the dressed states here are of
limited use since the nonadiabatic coupling is always significant. This is
related to the fact that the decay constants are intimately related to the
coupling strength, that the potentials periodically approach one another, and
that the nonadiabatic coupling is maximal at these close separations
[$\partial\theta/\partial x\sim (\sigma/d)\sin x]$. The dressed picture gives a
reasonable approximation to the friction and diffusion coefficients when
$\left\vert \alpha\pm\mathrm{d}\right\vert \gg1$ and $\mathrm{d}\gtrsim
\sigma\gg1$. In this limit one can make a secular approximation and ignore the
contribution from the $C(x)$ and $D(x)$ terms in Eqs. (\ref{dcf}). The
nonadiabatic terms neglected in Eq. (\ref{drs}) are of order $\sigma
\alpha/\mathrm{d}^{2}$ in this limit. Thus, the approximation is valid for
relatively large detunings and values of $\alpha$ less than or on the order of
unity. Indeed, in the limit $\sigma/\mathrm{d}\ll1$, the dressed picture
results reproduce those of Eq. (\ref{cs}) provided $\alpha$ is not too large.
On the other hand, they do not reproduce those of Eq. (\ref{cd}) near the
Doppler shifted two-photon resonances; the dressed results vary as $\left(
1+f^{2}\right)^{-1}$ rather than $\left(1+f^{2}+\sigma^{2}\right)^{-1}$.
Both the secular approximation and the neglect of nonadiabatic coupling break
down near these two-photon resonances.

For the nonadiabatic coupling to be negligible compared with convective
derivatives such as $\alpha\partial w_{D}/\partial x$, it is necessary that
$\partial\theta/\partial x\ll1$. It can be shown that in the regions of
closest approach of the potentials that $\partial\theta/\partial x\sim
\chi_{eff}/\delta=\sigma/\mathrm{d}$. Thus, for the dressed picture to be
valid, one is necessarily in the region where the approximate solutions
Eqs.(\ref{cs}), (\ref{cd}) are all that is needed.

\subsection{Density matrix solution}

As a final approximate approach one can adiabatically eliminate $\rho_{12}$
and $\rho_{21}$ from Eqs.(\ref{dens}). This procedure will allow one to
obtain an analytical solution for all density matrix elements in terms of a
sum over Bessel functions. Such an approach is valid for $\delta\gg
\Gamma^{\prime}$ \textit{and }$\delta\gg\chi^{2}/\Delta$ so it has a limited
range of applicability. The detailed results are not presented here.

\section{Momentum and energy distributions}
\begin{figure}[tH]
\centerline{\scalebox{0.5}{\includegraphics{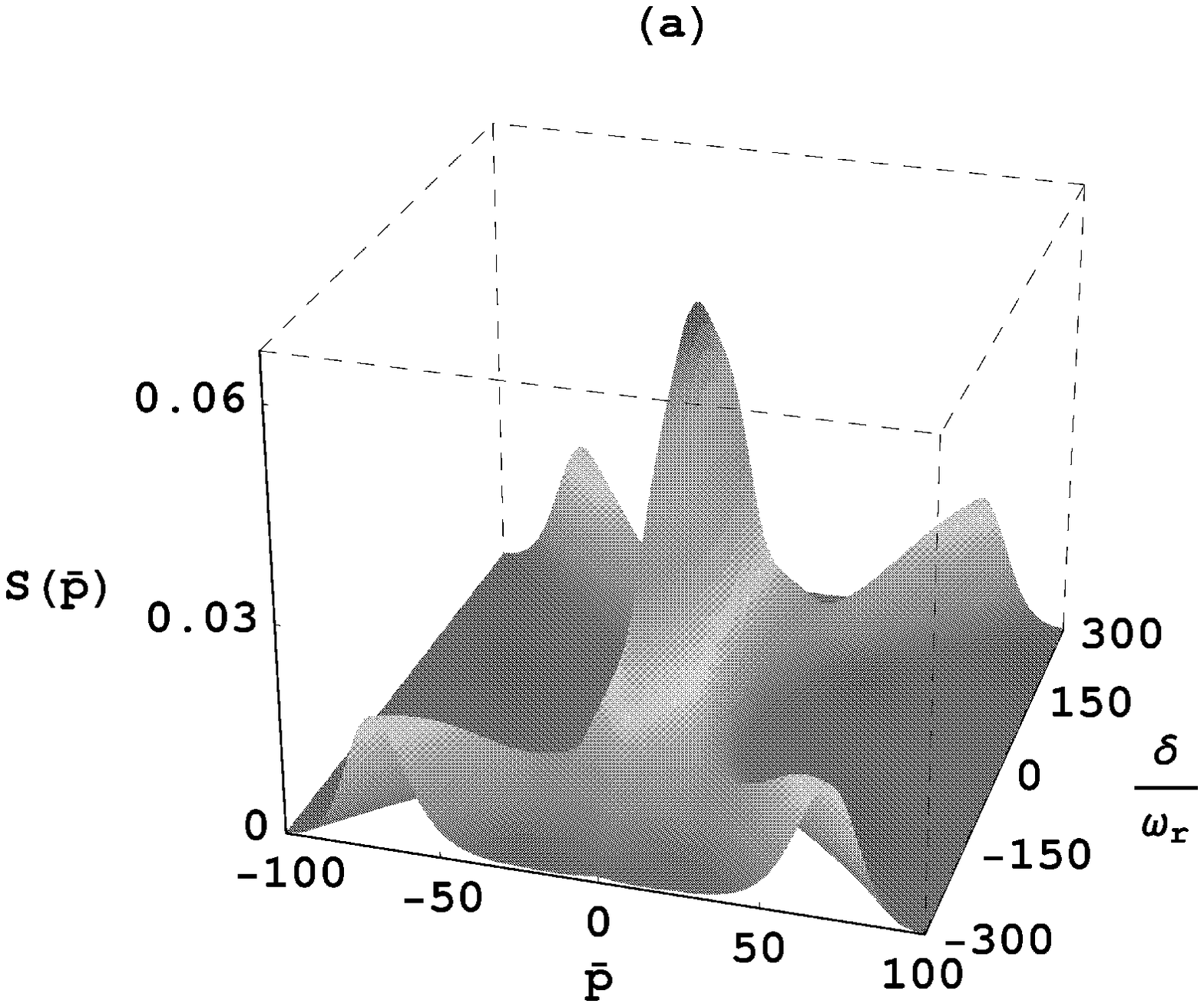}}
\scalebox{0.5}{\includegraphics{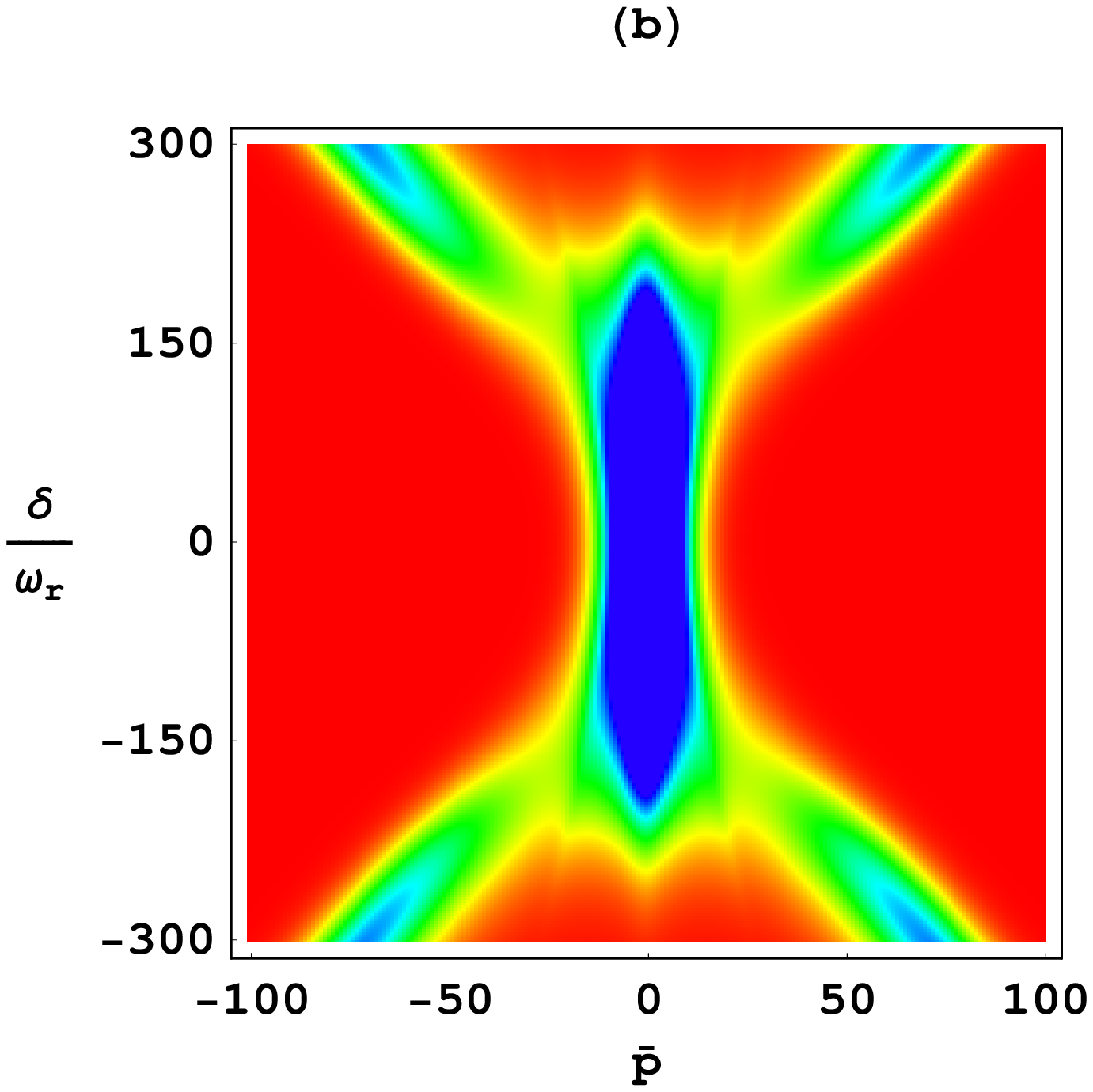}}} \caption{The momentum
distribution $S(\bar{p})$ as a function of
$\tilde{\delta}=\delta/\omega_{r}$ for $I=40$, $\sigma=10$;
3-dimensional plot (a) and density plot (b). }
\end{figure}

In terms of the normalized momentum $\bar{p}=p/\hbar k$, the steady state
solution of the Fokker-Planck equation,
\begin{equation}
\frac{\bar{D}_{tot}}{\hbar k}\frac{\partial S}{\partial\bar{p}}=\left(
\bar{F}\,-\frac{1}{\hbar k}\frac{\partial\bar{D}_{sp}}{\partial\bar{p}
}\right)  S\,, \label{eq}
\end{equation}
subject to the boundary condition $\partial S/\partial p|_{p=0}=0$, is
\begin{equation}
S(\bar{p})=S_{0}\exp\left\{\hbar k\int_{0}^{\bar{p}}d\bar{p}^{\prime}
\frac{\left(\bar{F}\,-\frac{1}{\hbar k}\frac{\partial\bar{D}_{sp}}
{\partial p^{\prime}}\right)  }{\bar{D}_{tot}}\right\}\,, \label{eqsol}
\end{equation}
where
\begin{equation}
S_{0}=\left[  \int_{-\infty}^{\infty}d\bar{p}\,\exp\left\{\hbar k
\int_{0}^{\bar{p}} d\bar{p}^{\prime}\frac{\left( \bar{F}\,-\frac{1}
{\hbar k}\frac{\partial \bar{D}_{sp}}{\partial\bar{p}^{\prime}}\right) }
{\bar{D}_{tot}}\right\}\right]^{-1}\,. \label{so}
\end{equation}
Taking into account definitions, Eq.(\ref{avg}), we obtain
\begin{equation}
S(\bar{p})=S_{0}\exp\left\{-\int_{0}^{\bar{p}}d\bar{p}^{\prime}
\frac{\displaystyle 2\sigma\xi_{f}}{\displaystyle \frac{7}{5}+
\frac{3}{10}\xi_{_{sp}}+ 4\sigma^{2}\xi_{st}}\right\}\,, \label{dist}
\end{equation}
where the $\xi$s are define in Eqs.(A17) and we neglect the term
$\left(  1/\hbar k\right)  \partial\bar{D}_{sp}/\partial\bar{p}^{\prime}$.
\begin{figure}[tH]
\centerline{\scalebox{0.5}{\includegraphics{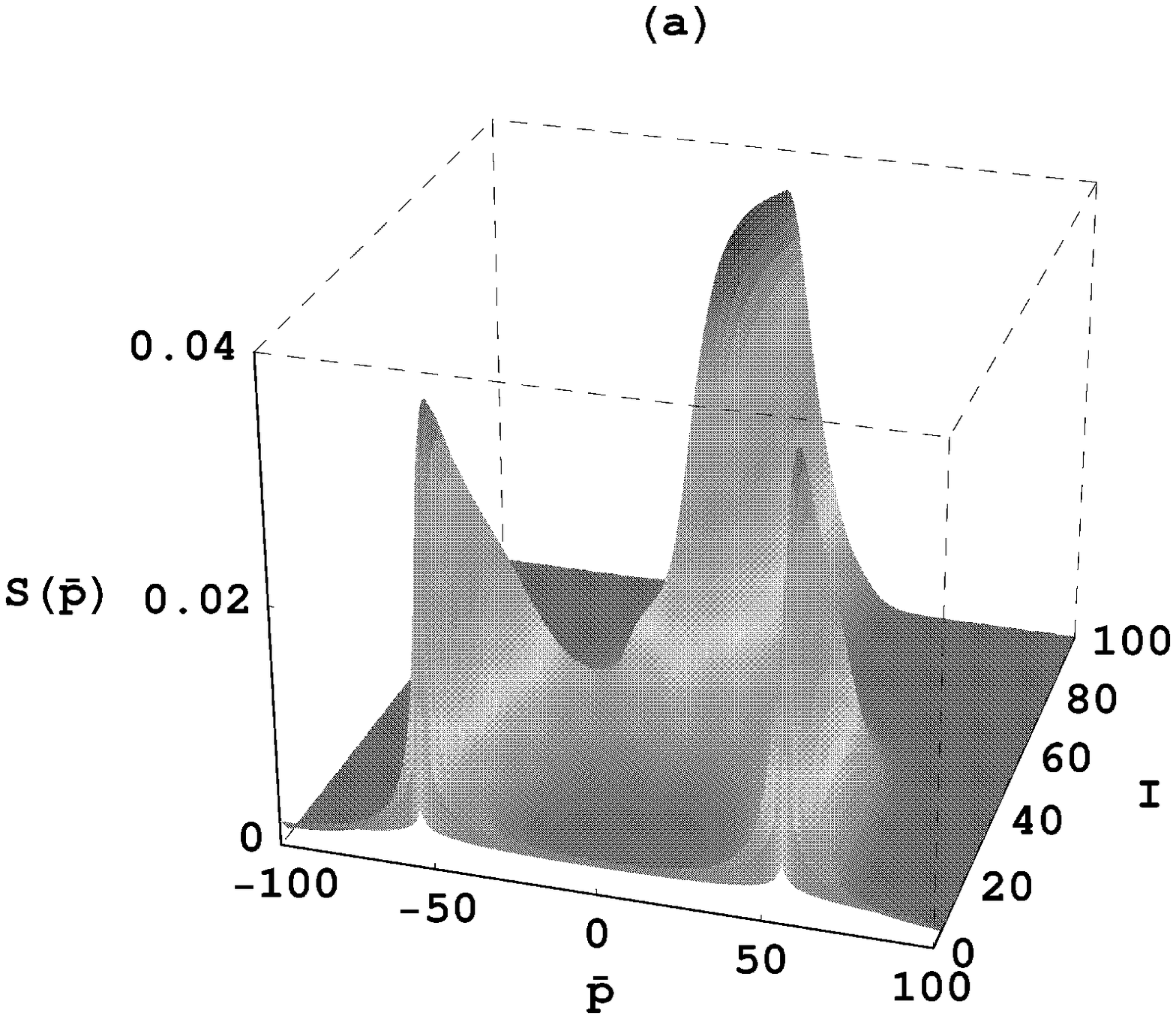}}
\scalebox{0.5}{\includegraphics{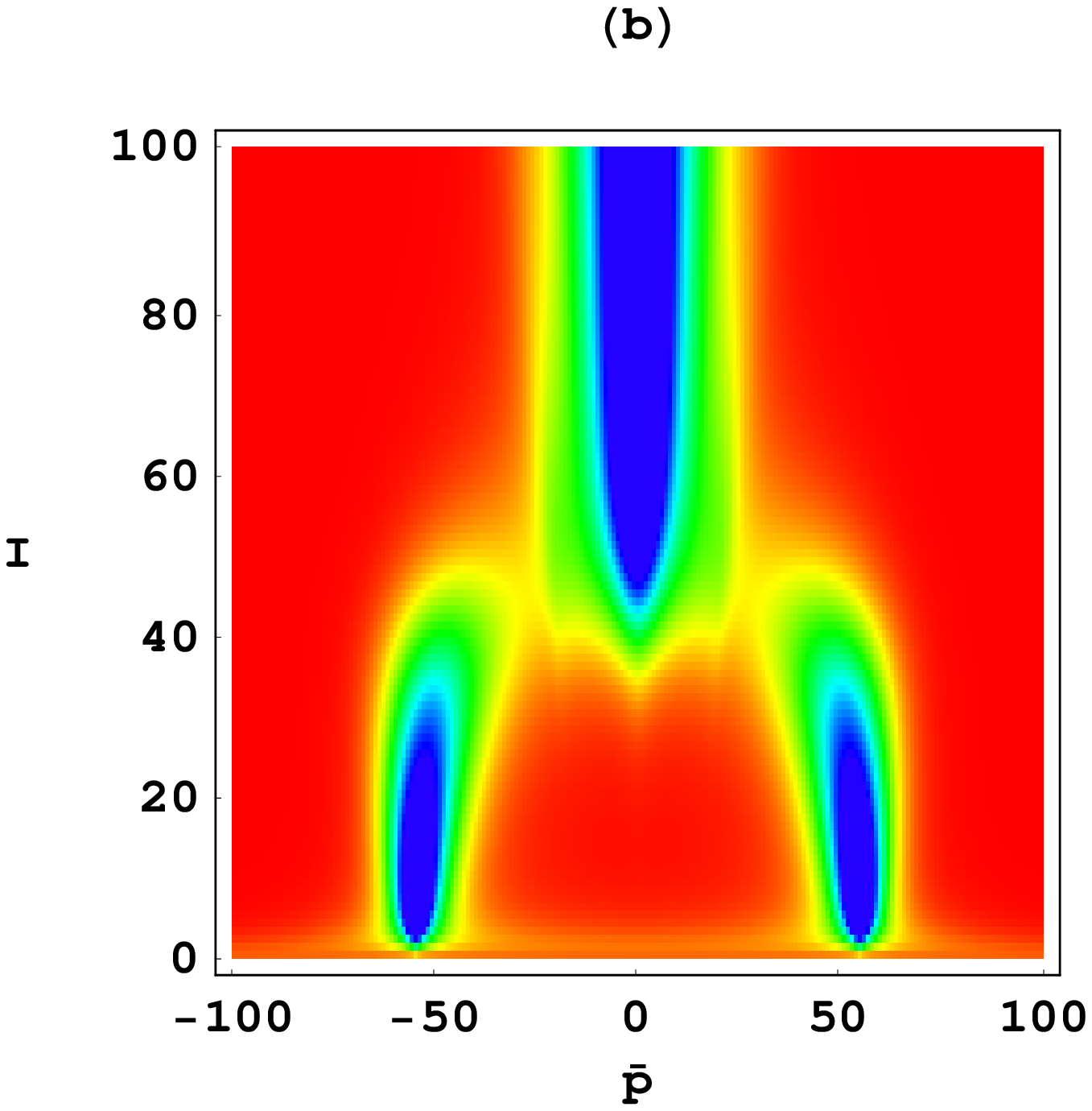}}} \caption{The momentum
distribution $S(\bar{p})$ as a function of $I$ for
$\tilde{\delta}=220$, $\sigma=10$;
3-dimensional plot (a) and density plot (b).} 
\end{figure}

The momentum distribution $S(\bar{p})$ is plotted in Fig. 5 as a function of
$\tilde{\delta}$ for $\sigma=10$ and $I=40$ and in Fig. 6 as a function of
$I$ for $\sigma=10$ and $\tilde{\delta}=220$. The curves in
Fig. 7 are cuts of Fig. 5 for $\tilde{\delta}=0$, $170$, $220$, and
$300$. When $\mathrm{d}/\sigma=\tilde{\delta}/2I\ll1$ and $I\gg\sigma>1$,
there is a central component having width of order $\sqrt{2I}\left(
1+\tilde{\delta}^{2}/8I^{2}\right)^{1/2}$, that is estimated using Eqs.
(\ref{ps}). For $\mathrm{d}/\sigma=\tilde{\delta}/2I\gg1$, the momentum
distribution breaks into three components centered at $\bar{p}=0,\pm
\tilde{\delta}/4$ $\left[2k\mathrm{v=0,}\pm\delta\right]$, with the
central component negligibly small compared with the side peaks \{relative
strength of side to central peak scales roughly as $\left( \mathrm{d}
/\sigma\right)^{5I/14}$, estimated using Eqs. (\ref{cs}), (\ref{cd})$\}$. The
width of the side peaks for $\mathrm{d}/\sigma=\tilde{\delta}/2I\gg1$ also
scale as $\sqrt{I}$, although they are slightly broader than the central peak
when $\mathrm{d}/\sigma=\tilde{\delta}/2I\ll1$, reflecting the fact that the
side peak cooling is of the corkscrew polarization nature, while the central
component cooling for $\mathrm{d}/\sigma\ll1$ is of the Sisyphus nature. For
intermediate values of $\mathrm{d}/\sigma$ three peaks in the momentum
distribution are seen clearly; for example, when $\tilde{\delta}
=220$, $I=40$, $\sigma=10$, the amplitudes of the three peaks are equal.
\begin{figure}[t]
\centerline{\scalebox{0.4}{\includegraphics{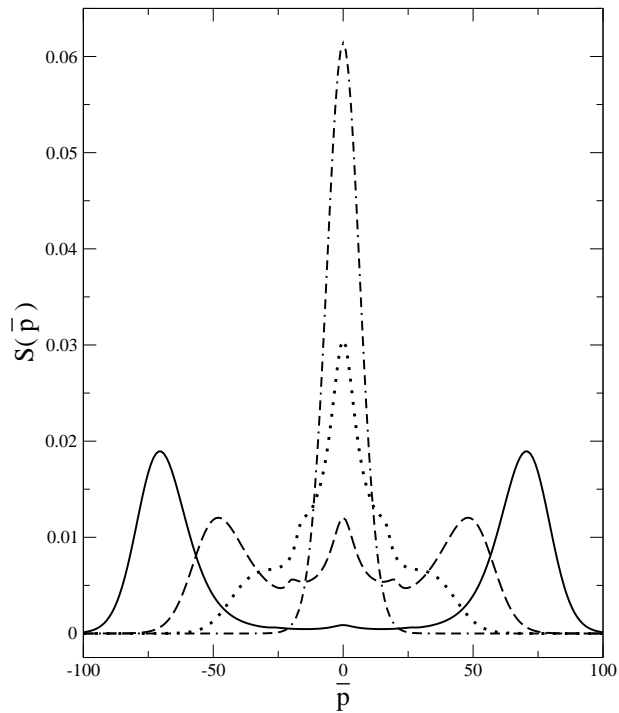}}}
\caption{Cuts in the momentum distribution of Fig.5 for
$\tilde{\delta }=0$(dash-dotted line), 170(dotted line),
220(dashed line), 300(solid line).} \label{cuts}
\end{figure}

The mean equilibrium kinetic energy can be calculated according to
\begin{equation}
\tilde{E}=E_{eq}/E_{r}=\int\limits_{-\infty}^{\infty}d\bar{p}\,\bar{p}
^{2}S(\bar{p})\,, \label{eqt}
\end{equation}
where $E_{r}=\hbar\omega_{r}$ is the recoil energy. This quantity must be
calculated numerically, in general. However, for $\mathrm{d}/\sigma
=\tilde{\delta}/2I\ll1$ and $I\gg\sigma>1$, one can estimate that $\tilde
{E}=I\left(1+\tilde{\delta}^{2}/8I\right)^{1/2}$, using Eqs.(\ref{ps}).
For $\mathrm{d}/\sigma=\tilde{\delta}/2I\gg1$, the side peaks lead to an
equilibrium energy that scales as $\left(\tilde{\delta}/4\right)^{2}$
since momentum components at both $\bar{p}=\pm\tilde{\delta}/4$ are present;
however, the energy width associated with each side peak scales as $I$. In
Fig.8, we plot $\tilde{E}=E_{eq}/E_{r}$ as a function of $I$ for $\sigma=10$
and several values of $\tilde{\delta}$.
\begin{figure}[t]
\vspace{1.5cm}
\centerline{\scalebox{0.4}{\includegraphics{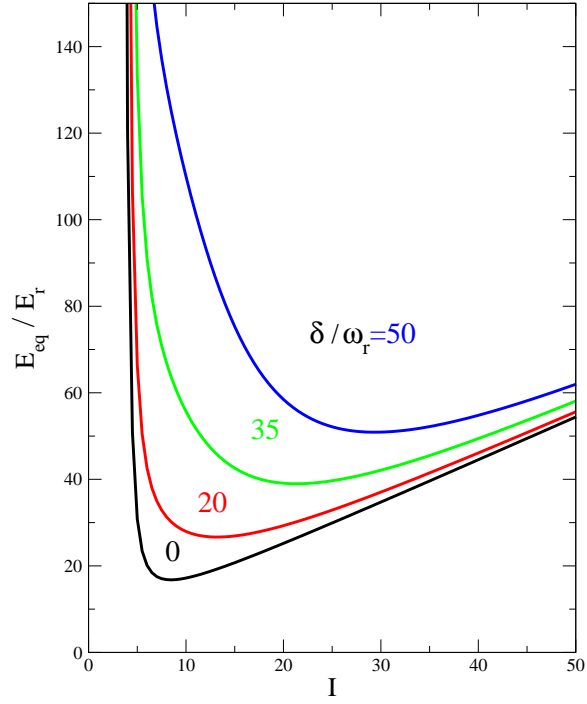}}}
\caption{Equilibrium energy in units of the recoil
energy as a function of dimensionless intensity $I$ for
$\sigma=10$ and $\tilde{\delta}=0,20,35,50$.} 
\end{figure}

\section{Summary}
We have extended the calculations of I to allow for non-zero detuning in a
standing-wave Raman scheme (SWRS) that results in reduced period optical
lattices. The results differ from that of conventional Sisyphus cooling.
Optimal cooling occurs for exact two-photon resonance, but many new and
interesting features appear for non-zero detuning. A dressed atom picture was
introduced, but had limited usefulness, owing to the presence of nonadiabatic
transitions. In a future planned publication, we will look at Monte Carlo
solutions to this problem and examine the dynamics of the velocity
distribution. Specifically we will attempt to determine how the atomic
momentum jumps between the momentum peaks shown in Fig. 5. Furthermore we will
see if it is possible to localize atoms in the potential wells shown in Fig.
4. The ability to do so would imply separation of $\lambda/8$
between atoms in adjacent wells.

\section{Acknowledgments}
This research is supported by National Science Foundation under Grants
No.\ PHY-0244841, PHY-0245522, and the FOCUS Center Grant. We thank G. Raithel
and G. Nienhuis for helpful discussions.

\section*{Appendix}
Using the Fourier series expansion
\begin{equation}\label{fs}
w(x)=\sum_{n}W_{n}e^{i2nx}\,, \tag{A1}
\end{equation}
in Eq. (\ref{weq1}) we obtain the recursion relation
\begin{equation}\label{rec}
\begin{array}[c]{l}
A_{-}(n)W_{n-1}+A_{0}(n)W_{n}+A_{+}(n)W_{n+1}=\\
S(p)\left[  B_{-}\delta_{n,-1}+B_{0}\delta_{n,0}+B_{+}\delta_{n,1}\right] \\
+2\hbar k\sigma\frac{\partial S}{\partial p}\left[  B_{-}^{\prime}
\delta_{n,-1}+B_{0}^{\prime}\delta_{n,0}+B_{+}^{\prime}\delta_{n,1}\right]\, ,
\end{array}
\tag{A2}
\end{equation}
where
\begin{subequations}
\label{a}
\begin{align}
A_{-}(n)  &  =\frac{{\sigma^{2}}}{{1+2in\alpha}}\frac{{1+i\alpha(2n-1)}
}{{\mathrm{d}^{2}+(1+i\alpha(2n-1))^{2}}}\,,\tag{A3a}\label{am}\\
A_{0}(n)  &  =1+\frac{{\sigma^{2}}}{{1+2in\alpha}}\left[  \frac{{1+i\alpha
(2n-1)}}{{\mathrm{d}^{2}+(1+i\alpha(2n-1))^{2}}}+\frac{{1+i\alpha(2n+1)}
}{{\mathrm{d}^{2}+(1+i\alpha(2n+1))^{2}}}\right]  \,,\tag{A3b}\label{a0}\\
A_{+}(n)  &  =\frac{{\sigma^{2}}}{{1+2in\alpha}}\frac{{1+i\alpha(2n+1)}
}{{\mathrm{d}^{2}+(1+i\alpha(2n+1))^{2}}}\,, \tag{A3c}\label{ap}
\end{align}
\end{subequations}
\begin{subequations}
\label{b}
\begin{align}
B_{-}  &  =\frac{{\sigma}}{{2}}\frac{{1}}{{1-i2\alpha}}\frac{{\mathrm{d}}
}{{\mathrm{d}^{2}+(1-i\alpha)^{2}}}\,,\tag{A4a}\label{bp}\\
B_{0}  &  =\frac{{\sigma}}{{2}}\left\{  \frac{{\mathrm{d}}}{{\mathrm{d}
^{2}+(1-i\alpha)^{2}}}+\frac{{\mathrm{d}}}{{\mathrm{d}^{2}+(1+i\alpha)^{2}}
}\right\}  \,,\tag{A4b}\label{bo}\\
B_{+}  &  =\frac{{\sigma}}{{2}}\frac{{1}}{{1+i2\alpha}}\frac{{\mathrm{d}}
}{{\mathrm{d}^{2}+(1+i\alpha)^{2}}}\,, \tag{A4c}\label{bm}
\end{align}
\end{subequations}
\begin{subequations}
\begin{align}
B_{-}^{\prime}  &  =\frac{{i\sigma}}{2}\frac{1}{{1-i2\alpha}}\frac
{{\mathrm{d}}}{{\mathrm{d}^{2}+(1-i\alpha)^{2}}}\,,\tag{A5a}\label{bpp}\\
B_{0}^{\prime}  &  =\frac{{i\sigma}}{2}\left\{  \frac{{\mathrm{d}}
}{{\mathrm{d}^{2}+(1-i\alpha)^{2}}}-\frac{{\mathrm{d}}}{{\mathrm{d}
^{2}+(1+i\alpha)^{2}}}\right\}  \,,\tag{A5b}\label{bp0}\\
B_{+}^{\prime}  &  =-\frac{{i\sigma}}{2}\frac{{1}}{{1+i2\alpha}}
\frac{{\mathrm{d}}}{{\mathrm{d}^{2}+(1+i\alpha)^{2}}}\,. \tag{A5c}\label{bpm}
\end{align}

We are faced with solving the following equation:
\end{subequations}
\begin{equation}
\left(
\begin{array}
[c]{ccccccc}
\ddots & \ddots & 0 & 0 & 0 & 0 & 0\\
\ddots & A_{0}(-2) & A_{+}(-2) & 0 & 0 & 0 & 0\\
0 & A_{-}(-1) & A_{0}(-1) & A_{+}(-1) & 0 & 0 & 0\\
0 & 0 & A_{-}(0) & A_{0}(0) & A_{+}(0) & 0 & 0\\
0 & 0 & 0 & A_{-}(1) & A_{0}(1) & A_{+}(1) & 0\\
0 & 0 & 0 & 0 & A_{-}(2) & A_{0}(2) & \ddots\\
0 & 0 & 0 & 0 & 0 & \ddots & \ddots
\end{array}
\right)  \left(
\begin{array}[c]{c}
\vdots\\
W_{-2}\\
W_{-1}\\
W_{0}\\
W_{1}\\
W_{2}\\
\vdots
\end{array}
\right)  =\left(
\begin{array}[c]{c}
\vdots\\
0\\
\tilde{B}_{-}\\
\tilde{B}_{0}\\
\tilde{B}_{+}\\
0\\
\vdots
\end{array}
\right)  , \tag{A6}\label{mrec}
\end{equation}
where
\begin{equation}
\tilde{B}_{j}=S(p)B_{j}+2\hbar k\sigma\frac{\partial S}{\partial p}
B_{j}^{\prime}\, , \tag{A7}\label{bt}
\end{equation}
and $j=\pm,0$. From Eq. (\ref{mrec}), we see that for $n<-1$ and for $n>1$
\begin{equation}
A_{-}(n)W_{n-1}+A_{0}(n)W_{n}+A_{+}(n)W_{n+1}=0. \tag{A8}\label{recre}
\end{equation}
The final solution for the spatially averaged friction and diffusion
coefficients depends only on $W_{0}$, $W_{\pm1}$. However, to calculate these
quantities all the other $W$s must be evaluated. In practice, we truncate Eq.
(\ref{mrec}) by setting $W_{\pm n}=0$ and then compare the solution with that
obtained by setting $W_{\pm\left(  n+1\right)  }=0$; when these solutions
differ by less than a fraction of a percent, we use the result to evaluate
$W_{\pm2}/W_{\pm1}$, from which one can then calculate $W_{0}$, $W_{\pm1}$.

For $n>1,$ Eq. (\ref{mrec}) yields
\[
W_{n}\left(  1+\frac{A_{+}(n)}{A_{0}(n)}\frac{W_{n+1}}{W_{n}}\right)
=-\frac{A_{-}(n)}{A_{0}(n)}W_{n-1} \, ,
\]
which can be written in the form
\[
\frac{W_{n}}{W_{n-1}}=-\frac{A_{-}(n)/A_{0}(n)}{{1}+\frac{{A_{+}(n)}}
{{A_{0}(n)}}\frac{{W_{n+1}}}{{W_{n}}}}.
\]
Setting $n=2$ we obtain the continued fraction solution
\begin{equation}
C_{+}=\frac{W_{2}}{W_{1}}=-\frac{\mathstrut A_{-}(2)/A_{0}(2)}{{1}
-\frac{\mathstrut{A_{+}(2)}}{{A_{0}(2)}}\frac{\mathstrut{A_{-}(3)/A_{0}(3)}
}{{1}-\frac{\mathstrut{A_{+}(3)}}{{A_{0}(3)}}\frac{\mathstrut{A_{-}
(4)/A_{0}(4)}}{{1}-{\hdots}}}} \, .\tag{A9}\label{cplus}
\end{equation}
Similarly, for $n<-1$ we find
\begin{equation}
C_{-}=\frac{W_{-2}}{W_{-1}}=-\frac{A_{+}(-2)/A_{0}(-2)}{{1}-\frac{{A_{-}(-2)}
}{{A_{0}(-2)}}\frac{{A_{+}(-3)/A_{0}(-3)}}{{1}-\frac{{A_{-}(-3)}}{{A_{0}(-3)}
}\frac{{A_{+}(-4)/A_{0}(-4)}}{{1}-{\hdots}}}} \, .\tag{A10}\label{cminus}
\end{equation}

One can now use Eqs. (\ref{mrec}), (\ref{cplus}), (\ref{cminus}) to obtain
equations for $W_{0}$, $W_{\pm1}$ in terms of $C_{\pm}$ and the
$\tilde{B}_{j}$s. Explicitly, one finds
\begin{equation}
\left(
\begin{array}
[c]{ccc}
A_{0}(-1)+A_{-}(-1)C_{-}\, & \,A_{+}(-1)\, & 0\\
A_{-}(0) & A_{0}(0) & A_{+}(0)\\
0 & \,A_{-}(1)\, & \,A_{0}(1)+A_{+}(1)C_{+}
\end{array}
\right)  \left(
\begin{array}[c]{c}
W_{-1}\\
W_{0}\\
W_{1}
\end{array}
\right)  =\left(
\begin{array}[c]{c}
\tilde{B}_{-}\\
\tilde{B}_{0}\\
\tilde{B}_{+}
\end{array}
\right)  . \tag{A11}\label{finW}
\end{equation}
The procedure is to obtain $C_{+}$ and $C_{-}$ according to the continued
fraction solutions Eq.~(\ref{cplus}) and Eq. (\ref{cminus}) and then find
$W_{-1,0,1}$ from Eq.~(\ref{finW}).

Next we calculate $\eta_{1,2}$, Eq.(\ref{eta1})-(\ref{eta2}) using Eqs.
(\ref{fs}), (\ref{finW}), (\ref{inta}) as
\begin{subequations}
\label{etffap}
\begin{align}
\eta_{1}  &  =-\left[  a_{0}S+2a_{1}\hbar k\sigma\frac{\partial S}{\partial
p}+W_{0}a_{2}+W_{1}a_{3}+W_{-1}a_{4}\right]  /2\,,\tag{A12a}\label{etffapa}\\
\eta_{2}  &  =-\left[  b_{0}S+2b_{1}\hbar k\sigma\frac{\partial S}{\partial
p}+W_{0}b_{2}+W_{1}b_{3}+W_{-1}b_{4}\right]  /2\,, \tag{A12b}\label{etffapb}
\end{align}
where
\end{subequations}
\begin{subequations}
\label{aap}
\begin{align}
a_{0}  &  =\int_{0}^{\infty}d\tau\,e^{-\tau}\cos\left(  \mathrm{d}\tau\right)
\sin\left(  \alpha\tau\right)  =\frac{{\alpha(1-\mathrm{d}^{2}+\alpha^{2})}
}{{(1+(\mathrm{d}-\alpha)^{2})(1+(\mathrm{d}+\alpha)^{2})}}\,, \tag{A13a}
\label{aapa}\\
a_{1}  &  =\int_{0}^{\infty}d\tau\,e^{-\tau}\cos\left(  \mathrm{d}\tau\right)
\cos\left(  \alpha\tau\right)  =\frac{{(1+\mathrm{d}^{2}+\alpha^{2})}
}{{(1+(\mathrm{d}-\alpha)^{2})(1+(\mathrm{d}+\alpha)^{2})}}\,, \tag{A13b}
\label{aapb}\\
a_{2}  &  =2\sigma\int_{0}^{\infty}d\tau\,e^{-\tau}\sin\left(  \mathrm{d}
\tau\right)  \sin\left(  \alpha\tau\right)  =\frac{{4\sigma d\alpha}
}{{(1+(\mathrm{d}-\alpha)^{2})(1+(\mathrm{d}+\alpha)^{2})}}\,, \tag{A13c}
\label{aapc}\\
a_{3}  &  =i\sigma\int_{0}^{\infty}d\tau\,e^{-\tau}\cos\left(  \mathrm{d}
\tau\right)  \,e^{-i\alpha\tau}=\frac{{i\sigma}\mathrm{d}}{\mathrm{d}
{^{2}+(1+i\alpha)^{2}}}\,,\tag{A13d}\label{aapd}\\
a_{4}  &  =-i\sigma\int_{0}^{\infty}d\tau\,e^{-\tau}\cos\left(  \mathrm{d}
\tau\right)  \,e^{i\alpha\tau}=\frac{-{i\sigma}\mathrm{d}}{\mathrm{d}
{^{2}+(1-i\alpha)^{2}}}\,, \tag{A13e}\label{aapde}
\end{align}

\end{subequations}
\begin{subequations}
\label{bap}
\begin{align}
b_{0}  &  =a_{1}\,,\tag{A14a}\label{bapa}\\
b_{1}  &  =-a_{0}\,,\tag{A14b}\label{bapb}\\
b_{2}  &  =2\sigma\int_{0}^{\infty}d\tau\,e^{-\tau}\sin\left(  \mathrm{d}
\tau\right)  \cos\left(  \alpha\tau\right)  =\frac{{2\sigma\mathrm{d}
(1+\mathrm{d}^{2}-\alpha^{2})}}{{(1+(\mathrm{d}-\alpha)^{2})(1+(\mathrm{d}
+\alpha)^{2})}}\,,\tag{A14c}\label{bapc}\\
b_{3}  &  =\sigma\int_{0}^{\infty}d\tau\,e^{-\tau}\cos\left(  \mathrm{d}
\tau\right)  \,e^{-i\alpha\tau}=\frac{{\sigma}\mathrm{d}}{{\mathrm{d}
^{2}+(1+i\alpha)^{2}}}\,,\tag{A14d}\label{bapd}\\
b_{4}  &  =\sigma\int_{0}^{\infty}d\tau\,e^{-\tau}\cos\left(  \mathrm{d}
\tau\right)  \,e^{i\alpha\tau}=\frac{{\sigma}\mathrm{d}}{{\mathrm{d}
^{2}+(1-i\alpha)^{2}}}\,. \tag{A14e}\label{bape}
\end{align}
Since each $W_{i}$ contains a term proportional to $S$ and another term
proportional to $\frac{\partial S}{\partial p}$, one has
\end{subequations}
\begin{equation}
W_{i}=\theta_{i}S+2\hbar k\sigma\zeta_{i}\frac{\partial S}{\partial p}\,,
\tag{A15}\label{wi}
\end{equation}
and Eqs.(A12) can be written as
\begin{subequations}
\label{etfap}
\begin{align}
\eta_{1}  &  =-\left[  S(a_{0}+\theta_{0}a_{2}+\theta_{1}a_{3}+\theta
_{-1}a_{4})+2\hbar k\sigma\frac{\partial S}{\partial p}(a_{1}+\zeta_{0}
a_{2}+\zeta_{1}a_{3}+\zeta_{-1}a_{4})\right]  /2\,,\tag{A16a}\label{etfapa}\\
\eta_{2}  &  =-\left[  S(b_{0}+\theta_{0}b_{2}+\theta_{1}b_{3}+\theta
_{-1}b_{4})+2\hbar k\sigma\frac{\partial S}{\partial p}(b_{1}+\zeta_{0}
b_{2}+\zeta_{1}b_{3}+\zeta_{-1}b_{4})\right]  \,/2. \tag{A16b}\label{etfapb}
\end{align}
\end{subequations}

Using Eqs. (A16) with the Fokker-Plank equations (\ref{fp1}) and (\ref{fp3}),
one can identify the averaged force and diffusion coefficients appearing in
Eqs. (\ref{avg}) with
\begin{subequations}
\begin{align}
\xi_{f}  &  =(a_{0}+\theta_{0}a_{2}+\theta_{1}a_{3}+\theta_{-1}a_{4}),
\tag{A17a}\label{zeta}\\
\xi_{sp}  &  =(b_{0}+\theta_{0}b_{2}+\theta_{1}b_{3}+\theta_{-1}b_{4}),
\tag{A17b}\label{zetb}\\
\xi_{st}  &  =(a_{1}+\zeta_{0}a_{2}+\zeta_{1}a_{3}+\zeta_{-1}a_{4}).
\tag{A17c}\label{zetc}
\end{align}
\end{subequations}


\begin{thebibliography}{9}
\bibitem {rmn}P.\ R.\ Berman, G.~Raithel, R.~Zhang, and V.~S.~Malinovsky,
Phys.Rev. A 72 (2005) 033415 . This article contains several
additional references.

\bibitem {der}See, for example, A. Derevianko and C. C. Cannon, Phys. Rev. A
70 (2004) 062319.

\bibitem {single}This condition is necessary to neglect the effects of fields
$E_{1}$ acting on the 2-3 transition and $E_{2}$ acting on the 1-3 transition
with regards to light shifts and optical pumping; however, it is possible to
neglect the effect of fields $E_{2}$ and $E_{1}$ driving coherent transitions
between levels 1 and 2 (with $E_{2}$ acting on the 1-3 transition and $E_{1}$
acting on the 2-3 transition) under the much weaker condition that the optical
pumping rates be much smaller than $\omega_{21}$.

\bibitem {cond}The condition needed to neglect modulated Stark shifts
resulting from the combined action of fields $E_{1}$ and $E_{2}$ (or $E_{2}$
and $E_{4}$), as well as transitions between levels 1 and 2 resulting from
fields $E_{1}$ and $E_{2}$ (or $E_{3}$ and $E_{2}$) is $\left\vert \Omega
_{3}-\Omega_{1}\right\vert \ll\left\vert \chi\chi^{\prime}/\Delta\right\vert $
and $\left\vert \Omega_{4}-\Omega_{2}\right\vert \ll\left\vert \chi
\chi^{\prime}/\Delta\right\vert $, where $\chi$ is a Rabi frequency associated
with the 1-3 transition and $\chi^{\prime}$ is a Rabi frequency associated
with the 2-3 transition.

\bibitem {intrep}The interaction representation is one in which $\rho
_{12}^{normal}=\rho_{12}\,e^{i\left(  \Omega_{1}-\Omega_{2}\right)  t}
=\rho_{12}\,e^{i\left(  \Omega_{3}-\Omega_{4}\right)  t}$, where $\rho_{12}$
is the density matrix element in the field interaction representation.

\bibitem {thaler}M Goldstein and R. M. Thaler, Tables and Aids to Comp.
12 (1958) 18; \textit{ibid.} 13 (1959) 102.

\bibitem {zieg}J. Ziegler and P. R. Berman, Phys.Rev. A 16 (1977) 681.

\bibitem {zieg2}J. Ziegler and P. R. Berman, Phys.Rev. A 15 (1977) 2042.

\bibitem {sismag}S-Q. Shang, B.Sheehy, P van der Straten, and H. Metcalf,
Phys. Rev. Lett. 65 (1990) 317; P. Berman, Phys.\ Rev. A 43 (1991) 1470
\, [Note that Eq. (48) is valid only for $\Delta
/\Gamma\lesssim1$]; P.~van~der Straten, S-Q. Shang, B.Sheehy, H.
Metcalf, and G. Nienhuis, Phys.\ Rev. A 47 (1993) 4160.

\bibitem {cohen}J. Dalibard and C. Cohen-Tannoudji, J. Opt. Soc. B 6 (1989)
2023.

\bibitem {moddress}As was discussed in I, in the limit that $\delta\sim0$, it
is more convenient to introduce "dressed" states via the definitions
\begin{align*}
\left\vert A\right\rangle ^{\prime}  &  =\left(  \left\vert 1\right\rangle
+\left\vert 2\right\rangle \right)  /\sqrt{2}, \\
\left\vert B\right\rangle ^{\prime}  &  =\left(  -\left\vert 1\right\rangle
+\left\vert 2\right\rangle \right)  /\sqrt{2},
\end{align*}
with corresponding eigenvalues $\lambda_{\pm}=\pm2 \hbar \chi_{eff}\cos x$.
These potentials cross, as shown in Fig. 7 (b) of I, but there is no coupling
between the eigenstates. For $\delta\neq0$, we have not found a general
transformation that minimizes the nonadiabatic coupling.
\end{thebibliography}
\end{document}